\documentclass[sigplan,10pt]{acmart}

\usepackage{tikz}
\usepackage{graphicx} 
\usepackage{enumitem}
\usepackage[dvipsnames]{xcolor}
\usepackage{tikz}
\usetikzlibrary{arrows.meta}
\usetikzlibrary{calc}
\usepackage{tabularx}
\usepackage{array} 
\usepackage{diagbox} 
\usepackage{setspace}
\usepackage{epsfig}
\usepackage{amsmath}
\usepackage{float}
\usepackage{capt-of}
\usepackage{fvextra}   
\usepackage{mdframed}  
\usepackage[most]{tcolorbox}
\usetikzlibrary{decorations.pathreplacing}
\usepackage{xcolor}
\usepackage[htt]{hyphenat}

\newcommand{\hlline}{%
  \rlap{%
    \color{yellow!50}%
    \rule[-0.3\baselineskip]{\linewidth}{\baselineskip}%
  }%
}

\usepackage{makecell}
\usepackage{url}
\usepackage{subcaption}
\usepackage[ruled]{algorithm2e}
\usepackage{booktabs}
\usepackage{multirow}
\usepackage{diagbox}
\usepackage{textcomp}
\usepackage{parcolumns}
\tcbuselibrary{listings}

\definecolor{DeepOrange}{RGB}{255, 87, 34}
\definecolor{CodeBg}{RGB}{248,248,252}
\definecolor{CodeFrame}{RGB}{220,220,230}
\definecolor{Teal}{RGB}{0,128,128}
\lstdefinelanguage{json}{
  morestring=[b]",         
  showstringspaces=false,
  literate=
   *{0}{{{\color{Purple}0}}}{1}
    {1}{{{\color{Purple}1}}}{1}
    {2}{{{\color{Purple}2}}}{1}
    {3}{{{\color{Purple}3}}}{1}
    {4}{{{\color{Purple}4}}}{1}
    {5}{{{\color{Purple}5}}}{1}
    {6}{{{\color{Purple}6}}}{1}
    {7}{{{\color{Purple}7}}}{1}
    {8}{{{\color{Purple}8}}}{1}
    {9}{{{\color{Purple}9}}}{1}
}

\lstdefinestyle{CodeBase}{
  basicstyle=\ttfamily\scriptsize,
  backgroundcolor=\color{CodeBg},
  frame=single,
  rulecolor=\color{CodeFrame},
  numbers=left,
  numberstyle=\tiny\color{gray},
  numbersep=8pt,
  showstringspaces=false,
  tabsize=2,
  breaklines=true,
  breakatwhitespace=true,
  breakautoindent=true
  keepspaces=true,
  xleftmargin=.15in,
  framexleftmargin=.15in
}

\lstdefinestyle{CodeJson}{
  style=CodeBase,
  language=json,
  stringstyle=\color{RoyalBlue},   
  commentstyle=\color{Gray},
}

\lstdefinestyle{CodeCpp}{
  style=CodeBase,
  language=C++,
  keywordstyle=\color{RoyalBlue}\bfseries,
  commentstyle=\color{ForestGreen},
  stringstyle=\color{BrickRed},
  identifierstyle=\color{Black},
  morekeywords=[2]{std,size_t,uint32_t,uint64_t},
  keywordstyle=[2]\color{MidnightBlue}\bfseries,
  moredelim=[is][\color{BurntOrange}]{§}{§}, 
}

\lstdefinestyle{CodeCuda}{
  style=CodeBase,
  language=C++,
  keywordstyle=\color{Teal}\bfseries,
  commentstyle=\color{darkgray},
  stringstyle=\color{BurntOrange},
  identifierstyle=\color{Black},
  morekeywords={
    __global__,__device__,__host__,
    __shared__,__constant__,__syncthreads,
    dim3,blockIdx,threadIdx,blockDim,gridDim,
    TORCH_LIBRARY_EXPAND,scalar_t,TORCH_EXTENSION_NAME,torch::Tensor,at::cuda::OptionalCUDAGuard,device_of,cudaStream_t,at::cuda::getCurrentCUDAStream,std::enable_if_t,std::min,bfloat16,dim3,_typeConvert
  },
  keywordstyle=[2]\color{PineGreen}\bfseries,
}

\lstdefinestyle{CodePython}{
  style=CodeBase,
  language=Python,
  keywordstyle=\color{Orange}\bfseries,
  commentstyle=\color{Gray},
  stringstyle=\color{BrickRed},
  identifierstyle=\color{Black},
  morekeywords=[2]{self,True,False,None},
  keywordstyle=[2]\color{OliveGreen}\bfseries,
  literate=
    {input}{{\textcolor{Black}{input}}}{5}
    {apply}{{\textcolor{Black}{apply}}}{5},
}

\newcommand{\shihao}[1]{{\color{orange} \sf (SH: #1)}}

\newcommand{\mengting}[1]{{\color{brown} \sf (MT: #1)}}
\newcommand{\songlh}[1]{{\color{red} \sf (LH: #1)}}

\newcommand{\ignore}[1]{}

\newcommand{\indentbolditalicparagraphnospace}[1]{{\textit{\textbf{#1}}}}
\newcommand{\indentitalicparagraph}[1]{\underline{\textit{#1}}}

\newcommand{\boldunderparagraph}[1]{\vspace*{0.5ex}\noindent\underline{\textbf{#1}}}

\newcommand{\beforecaption}{\vspace{-.15cm}\begin{spacing}{0.85}}
\newcommand{\aftercaption}{\vspace{-.15cm}\end{spacing}}

\newcommand{\mycaption}[3]{\beforecaption\caption{\label{#1}{#2} #3}\aftercaption}

\newcommand{\eg}{\textit{e.g.}}

\newcommand{\etal}{\textit{et al.}}

\newcolumntype{Y}{>{\centering\arraybackslash}X}

\definecolor{promptbg}{HTML}{F8F9FF}
\definecolor{promptborder}{HTML}{DDE3FF}

\newcommand{\Tool}{M2K\xspace}

\newcommand{\DTool}{HFProbe\xspace}
\newcommand{\STool}{cuKLEE\xspace}

\copyrightyear{2026}
\acmYear{2026}
\setcopyright{cc}
\setcctype{by}
\acmConference[SOSP '26]{ACM SIGOPS 32nd Symposium on Operating Systems Principles}{September 29-October 02, 2026}{Prague, Czech Republic}
\acmBooktitle{ACM SIGOPS 32nd Symposium on Operating Systems Principles (SOSP '26), September 29-October 02, 2026, Prague, Czech Republic}
\acmDOI{10.1145/3830418.3843908}
\acmISBN{979-8-4007-2585-2/2026/09}

\begin{document}

\title{\Tool{}: Making the Model-Kernel Interface Explicit for Reliable CUDA Kernel Verification}


\author{Mengting He}
\affiliation{
  \institution{Penn State}
  \country{USA}
}

\author{Shihao Xia}
\affiliation{
  \institution{Penn State}
  \country{USA}
}

\author{Haomin Jia}
\affiliation{
  \institution{SKLP, ICT, CAS}
  \country{China}
}

\author{Wenfei Wu}
\affiliation{
  \institution{Peking University}
  \country{China}
}

\author{Linhai Song}
\affiliation{
  \institution{SKLP, ICT, CAS}
  \country{China}
}

\begin{abstract}

Large language model (LLM) inference systems rely on CUDA kernels for core GPU computations, 
yet the interface between models and kernels is implicit and poorly specified. 
Models and kernels evolve independently and often make 
incompatible assumptions about tensor shapes and input sizes, 
leading to subtle memory bugs in CUDA kernels.
These bugs can crash inference services, corrupt model weights,
or be exploited by remote adversaries. 
Existing techniques either incur prohibitive runtime overhead, require specialized hardware, 
or fail to handle dynamic tensor shapes and variable kernel launch configurations, leaving the CUDA memory bugs largely unaddressed. 

This paper presents \Tool{}, a fully automated framework that makes the 
model-kernel interface explicit and leverages it to detect memory bugs 
in CUDA kernels used in LLM inference systems.
\Tool{} consists of two components.
\DTool{} traces model execution without GPU hardware, 
classifies kernel arguments into model-fixed and user-variable, 
and emits symbolic constraints that capture the interface. 
\STool{} then performs symbolic execution 
on CUDA kernels to pinpoint memory bugs under 
the interface constraints, 
modeling tensors as disjoint memory regions and treating thread identifiers symbolically to scale to thousands of threads.
In the evaluation, \Tool{} discovers 181 previously unknown bugs 
in real LLM inference systems, while producing only nine false positives, demonstrating its effectiveness.

\end{abstract}


\settopmatter{printfolios=false, authorsperrow=5}
\maketitle
\pagestyle{empty}

\section{Introduction}
\label{sec:intro}

The rapid adoption of large language models (LLMs) has driven 
an explosion of inference workloads across chatbots~\cite{adamopoulou2020chatbots, adamopoulou2020overview, dam2024complete}, search~\cite{lewis2020retrieval, zhu2025large}, 
recommendation~\cite{wu2024survey, zhao2024recommender, bao2023large}, 
copilots~\cite{chen2021evaluating, zheng2023survey, jiang2024survey}, scientific computing~\cite{wang2023scientific, lewkowycz2022solving, boiko2023emergent}, 
and countless other areas~\cite{zhao2023survey, bommasani2021opportunities, bubeck2023sparks}. 
These services now operate at scale and serve real users, 
making LLM inference systems a critical part of modern computing infrastructure.

LLM inference systems rely on a layered software architecture. Model developers publish Python model files and weights (via Hugging Face~\cite{huggingface}), 
which inference frameworks (\eg, vLLM~\cite{vLLM}, SGLang~\cite{sglang}, 
Transformers~\cite{transformers}) load and execute. 
These frameworks manage inference requests and 
map high-level tensor operations in Python model code
to host-side wrapper code. 
The wrapper code prepares CUDA 
kernel inputs and configures kernel launches. At the lowest layer, CUDA kernels perform the core linear algebra 
computation on the GPU. 
This separation enables independent development at each layer and accelerates system evolution.

However, the layered architecture introduces a fragile, implicit interface between models and CUDA kernels.
The correctness of the system depends on assumptions made independently 
by the two sides, yet these assumptions are neither explicitly 
specified nor consistently enforced.
The mismatch manifests in two  ways. 
First, kernels may contain outright implementation errors in the intricate performance optimizations (\eg, vectorization, tiling),  
causing them to produce correct results 
only for specific tensor shapes, 
while silently performing out-of-bounds accesses for others. 
Second, kernels may impose implicit limits on input sizes 
(\eg, due to integer types used for index arithmetic or fixed-size internal buffers) 
that were safe when the kernel was written but are increasingly violated 
as models grow larger~\cite{zhao2023survey} and inference sequences become longer~\cite{team2024gemini, dao2022flashattention}. 
Both classes of mismatch lead to kernel memory bugs (\eg, buffer overflows, 
integer overflows) that are difficult to catch, because they surface only under specific combinations of model architecture, 
runtime input, and kernel implementation.

These kernel memory bugs carry severe consequences. 
Out-of-bounds accesses and data races trigger ``illegal memory access''
errors on the GPU, potentially crashing the entire inference service~\cite{park2021mind}. 
More critically, recent work has shown that memory bugs 
on GPUs can be exploited to modify arbitrary GPU memory 
or achieve arbitrary code execution~\cite{fun-profit,di2016study,miele2016buffer}. 
Since inference systems are exposed to the Internet and process user-controlled prompts, 
attackers can choose sequence length and influence server-side batching. 
As a result, exploitable memory bugs in CUDA kernels form a real attack surface that may allow attackers to compromise the host system or affect other service users, as demonstrated by our PoC exploits~\cite{TokenSmash, TokenBleed}.

Existing techniques for detecting memory bugs in CUDA kernels 
fall short in this setting.
Dynamic methods track metadata for GPU memory buffers and 
validate pointer dereferences at 
runtime~\cite{cuCatch,cudamemcheck,computesanitizer,lee2022securing,ziad2025gpuarmor,zhang2022lak,sullivan2023implicit,eizenberg_barracuda_2017,grace,gmrace,li2014ldetector,di2018gmod,di2020efficient}. 
However, they either incur high overhead 
or require hardware extensions not yet widely available. 
More fundamentally, they analyze kernels in isolation from the models that invoke them, leading to both false positives and false negatives.
Static tools~\cite{betts2012gpuverify,li2010symbolic,li2014practical,li2013se,li2012gklee,pereira2016verifying,li2019detecting,mai2023honeycomb,chatterjee2025proofwright,tong2024towards} 
cannot handle runtime-determined buffer sizes 
or the variability of kernel launch configurations, 
both of which are common in inference systems.
Techniques for testing machine learning frameworks~\cite{liu2023nnsmith, ou2025mirrorfuzz} 
mainly target the framework layer and cannot pinpoint CUDA kernel bugs.
In short, no existing technique bridges the gap between 
model-level usage and kernel-level implementation 
to provide comprehensive memory-bug detection for CUDA kernels.

We make a key observation: the implicit, unspecified interface between 
models and CUDA kernels is not merely a source of kernel memory bugs---it is 
an \emph{architectural weakness} of current LLM inference stacks. 
Models and kernels evolve on independent release cycles, 
maintained by different teams with different assumptions, 
yet no contract governs what one side may assume about the other. 
Effectively detecting memory-safety bugs in CUDA kernels 
therefore requires 
making this interface explicit: we must understand how each model 
invokes its kernels and use that information to systematically verify 
two properties---whether model usage respects kernel limitations, 
and whether kernel implementations are safe under all inputs the model can produce.

Realizing this idea raises two key challenges. 
First, \textit{how to automatically extract representative kernel usage patterns 
from a model.}
Kernel arguments are jointly determined by model architecture, 
runtime configuration, and user-controlled inputs 
(\eg, sequence length, batch size).
A practical solution must effectively characterize this combinatorial space of possible invocations
to provide actionable guidance for CUDA kernel analysis. 
Second, \textit{how to accurately and scalably verify CUDA kernels 
under the resulting usage conditions.}
Kernels operate on tensors whose shapes are determined at runtime, 
a setting that existing static analyses handle poorly. 
Moreover, kernels execute across thousands of parallel threads, 
and naïvely enumerating those threads causes state explosion,
making exhaustive verification infeasible.

\begin{figure}[t]
\centering
    \includegraphics[width=\columnwidth]{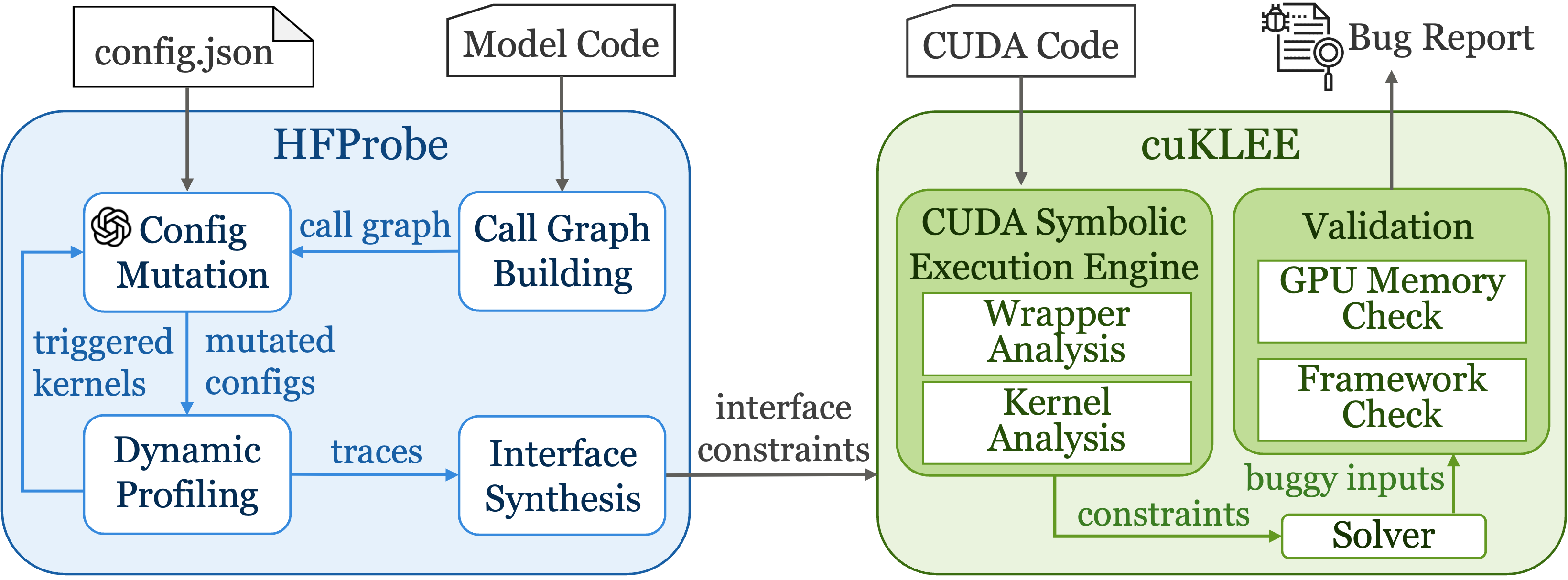} 
    \vspace{-0.15in}
    \mycaption{fig:overview}{An overview of \Tool{}.}
    {}
\vspace{-0.1in}
\end{figure}

To address these challenges, we design {\Tool{}}, 
a fully automated analysis framework that detects memory-safety issues 
in CUDA kernels as they are used by a given LLM model.
\Tool{} targets four types of bugs in CUDA kernels: 
buffer overflows, integer overflows, 
data races, and null-pointer dereferences, and supports the standard 
Hugging Face model format, covering the vast majority of publicly available LLMs.
As shown in Figure~\ref{fig:overview}, \Tool{} consists of two major components:  \DTool{} and \STool{}.

\DTool{} ``executes'' each model without requiring GPU hardware. 
It profiles kernel invocations by simulating execution without running the kernels, thereby achieving high efficiency.
It classifies kernel arguments into those fixed by the model architecture 
(\eg, hidden dimension) 
and those varying with user input (\eg, sequence length, batch size). 
\DTool{} translates this classification into symbolic constraints 
that make the model-kernel interface explicit 
and passes them to \STool{} for bug detection. 
In addition, \DTool{} statically identifies all CUDA kernels reachable from a given model and automatically mutates the model’s 
configurations to trigger more kernels, ensuring comprehensive analysis.

{\STool{}} performs symbolic execution on CUDA kernels with the interface constraints from \DTool{}. 
To handle input tensors whose memory layouts are determined at runtime, 
\STool{} models each tensor as a distinct, widely separated memory region. 
This abstraction simplifies pointer–tensor association and 
makes bounds checking more tractable. 
\STool{} introduces symbolic variables to represent each tensor’s base address, 
element count, dimensions, and other properties.
With these variables, \Tool{} can handle more than 100 tensor API methods by adding the appropriate constraints to its execution states.
It also models CUDA thread identifiers symbolically, 
allowing a single pass to capture both shared and thread-specific computations across thousands of threads.
Finally, \STool{} includes a memory usage model that 
accounts for the input token count, model size, and available GPU memory. 
It uses the model to determine whether a detected memory bug can actually 
occur on a target GPU, 
ensuring that reported bugs are actionable in practice.

\definecolor{ArrowOrange}{RGB}{255, 87, 34}

\begin{figure*}[t!]
\centering
\begin{minipage}[t]{0.48\textwidth}

\begin{tikzpicture}[remember picture, overlay, xshift=-0.3em, yshift=0.3em]

\node[anchor=north west] (codeA) at (0,0) {%
    \begin{minipage}{\textwidth}
\begin{lstlisting}[
            style=CodePython
        ]
class Qwen2ForCausalLM(nn.Module):
  def __init__(self, config):
    decoder_layer_type = Qwen2DecoderLayer
    self.layers = ...
  def forward(self, input_ids, positions):
    hidden_states, residual = ...
    for layer in self.layers:
      hidden_states, residual = layer(positions, hidden_states, residual)
    return hidden_states
class Qwen2DecoderLayer(nn.Module):
  def __init__(self, config):
    self.fp8_linear = Fp8LinearOp()
  def forward(self, positions, x, residual):
    out_dtype = torch.get_default_dtype()
    hidden_states = self.fp8_linear.apply(x, out_dtype)
    ...
    return hidden_states, residual
class Fp8LinearOp:
  def apply(self, input, out_dtype):
    in_2d = input.view(-1, input.shape[-1])
    qinput = torch.empty(in_2d.shape, dtype=out_dtype)
    scale = torch.zeros(1, dtype=torch.float32)
    torch.ops._C.dynamic_scaled_fp8_quant(qinput, in_2d, scale)
    return w8a8_scaled_mm_func(...)
\end{lstlisting}
    \end{minipage}
};


\draw[-{Latex[length=2mm]}, ArrowOrange, thick]
  ($(codeA.north west)+(16.5em,-7em)$)
    .. controls +(3em,-1em) and +(3em,-0.1em) ..
  ($(codeA.north west)+(16.9em,-11.5em)$);

\draw[-{Latex[length=2mm]}, ArrowOrange, thick]
  ($(codeA.north west)+(15em,-13.4em)$)
    .. controls +(3em,-1em) and +(3em,-0.3em) ..
  ($(codeA.north west)+(14.2em,-16.2em)$);

\end{tikzpicture}

\vspace{20.5em}
\subcaption{The simplified Python model code of Qwen2ForCausalLM.}
\label{fig:example-a}
\vspace{0.18em}
\begin{lstlisting}[
     style=CodeCpp
]
TORCH_LIBRARY_EXPAND(TORCH_EXTENSION_NAME, ops) {
  ops.def("dynamic_scaled_fp8_quant(Tensor! result, Tensor input, Tensor! scale) -> ()");
  ops.impl("dynamic_scaled_fp8_quant", torch::kCUDA, &dynamic_scaled_fp8_quant);
}
\end{lstlisting}
\vspace{-0.05in}
\subcaption{The registration of a CUDA kernel in Python.}
\label{fig:example-b}
\end{minipage}
\hfill
\begin{minipage}[t]{0.48\textwidth}
\begin{lstlisting}[
    style=CodeCpp
]
void dynamic_scaled_fp8_quant(torch::Tensor& out, torch::Tensor const& input, torch::Tensor& scale) {
  int64_t num_tokens = input.numel() / input.size(-1);
  int64_t num_elems = input.numel();
  dim3 grid(num_tokens);
  dim3 block(1024);
  using bf16 = at::BFloat16; using fp8 = at::Float8_e4m3fn;
  vllm::scaled_fp8_quant_kernel<bf16, fp8>
    <<<grid, block, 0>>>(out.data_ptr<fp8>(), input.data_ptr<bf16>(), scale.data_ptr<float>(), num_elems);
}
\end{lstlisting}
\vspace{-0.1in}
\subcaption{Wrapper function.}
\label{fig:example-c}

\vspace{0.2em}

\begin{lstlisting}[
    style=CodeCuda,
    escapechar=|,
    keepspaces=true,
]
template <typename scalar_t, typename fp8_type>
__global__ void scaled_fp8_quant_kernel(fp8_type* __restrict__ out, const scalar_t* __restrict__ input, const float* __restrict__ scale, int64_t num_elems) {
|{\hlline}\ | int tid = blockDim.x * blockIdx.x + threadIdx.x; // |{\color{red}overflow}|
  const float inverted_scale = 1.0f / (*scale);
  int step = blockDim.x * gridDim.x; // |{\color{red}overflow}|
  
  using float8x4_t = q8x4_t<fp8_type>;
  auto* v_in = reinterpret_cast<vec4_t<scalar_t> const*>(input);
  auto* v_out = reinterpret_cast<float8x4_t*>(out);
  int64_t const num_vec_elems = num_elems >> 2;
  
|{\hlline}\ |  for (int64_t i = tid; i < num_vec_elems; i += step) {
|{\hlline}\ |    vec4_t<scalar_t> in_vec = v_in[i]; // |{\color{red}OOB}|
|{\hlline}\ |    v_out[i] = scaled_fp8_conversion<fp8_type> // |{\color{red}OOB}|
|{\hlline}\ |      (in_vec, inverted_scale);
|{\hlline}\ | }
  ...
}


\end{lstlisting}
\vspace{-0.05in}
\subcaption{CUDA kernel. \textit{(Thread-dependent computations are in yellow.)}}
\label{fig:example-d}

\end{minipage}
\vspace{-0.05in}
\mycaption{fig:example}{A CUDA kernel example, its wrapper function, its registration to vLLM, and a Hugging Face model using the kernel.}
{}

\vspace{-0.05in}
\end{figure*}

We evaluate \Tool{} on CUDA kernels and models from three sources: all text-generation models in vLLM~\cite{vLLM} and the related kernels, 
all Hugging Face models with custom kernels, 
and models and kernels from four recent research publications~\cite{you2024shiftaddllm,malinovskii2024pvtuning,park2024anyprecision,huang2025mixturecompressor}.
In total, \Tool{} identifies 181 memory bugs, including 162 integer overflows 
\footnote{We suspect that integer overflows are prevalent in our detection results because they are more likely to cause silent data corruption than crashes.} 
and 19 out-of-bounds accesses, with only nine false positives, demonstrating
its effectiveness in detecting memory safety issues for CUDA kernels.
We have reported all 26 bugs detected in vLLM. So far, eight bugs
have been fixed in subsequent releases and additional four bugs have been confirmed by the programmers, underscoring \Tool{}'s practicality. 
%
We further compare \Tool{} with Honeycomb~\cite{mai2023honeycomb}, 
GKLEE~\cite{li2012gklee}, and 
ESBMC-GPU~\cite{pereira2016verifying} on 20 known vLLM CUDA kernel bugs.
\Tool{} successfully detects 15 of these bugs, far more than the baseline techniques, highlighting its superior bug coverage and 
clear improvement over existing techniques.
An ablation study shows that both \DTool{} and \STool{} are critical for \Tool{}'s effectiveness.
In sum, we make the following contributions:

\begin{itemize}[noitemsep, topsep=0pt, leftmargin=.25in]

\item We identify the implicit model-kernel interface 
as one source
 of 
memory bugs in CUDA kernels used in LLM inference systems 
and argue that making this interface explicit is the key 
to effective bug detection.
 
\item We develop \DTool{}, a dynamic model profiler that traces 
how models invoke CUDA kernels and infers model-kernel
interfaces, without  GPU hardware.
 
\item We design \STool{}, a symbolic execution engine specialized 
for CUDA kernels that supports tensors with dynamic shapes 
and scales to thousands of threads.
 
\item By integrating \DTool{} and \STool{}, we build \Tool{}, 
a fully automated framework for detecting CUDA memory bugs 
in LLM inference systems and demonstrate its effectiveness 
via thorough experiments.

\end{itemize}

 \Tool{} is publicly available at \url{https://github.com/system-pclub/M2K}.

\section{Background}
\label{sec:back}

This section provides background for the project, including 
Hugging Face models, the CUDA programming language, 
and CUDA kernel memory bugs in LLM inference systems. 

\subsection{Hugging Face Models}
\label{sec:hugging}

Hugging Face is the largest platform for 
sharing pre-trained models~\cite{huggingface}, 
offering both a model marketplace and libraries that facilitate model integration 
into LLM inference frameworks (\eg, vLLM~\cite{vLLM}, SGLang~\cite{sglang}, Transformers~\cite{transformers}). Today, it hosts millions of models, including major open-source foundation models (\eg, Qwen, LLaMA) and their variants, making it a central hub for the AI community.

A model released on Hugging Face typically contains 
four types of files. 
The \textit{config.json} file defines the 
model's architecture, including layer number, 
hidden size, activation functions, 
and other structural parameters. 
The model-weight files store the trained parameters. 
The Python code files define classes that extend \texttt{nn.Module} 
and implement the model architecture.
One top-level class assembles all model components, 
with its \texttt{forward()} method
as the entry of the model's whole forward pass.
For example, Figure~\ref{fig:example-a} shows simplified model code for Qwen2ForCausalLM, and \texttt{forward()} on line 5 begins the model’s inference process. 
Finally, documentation files (\eg, README.md) provide instructions, usage examples,
and licensing information.

\subsection{The CUDA Programming Language}
\label{sec:cuda}

CUDA is the dominant programming platform
for exploiting GPU computational power~\cite{cuda},
and most GPU kernels in inference systems are written in CUDA C++. 

CUDA source files typically contain two types 
of code: host wrapper functions and GPU kernel functions.
Host wrappers run on the CPU and are exposed to 
Python code in inference systems via registration or binding macros~\cite{TORCH_LIBRARY_EXPAND,PYBIND11_MODULE} (\eg, Figure~\ref{fig:example-b}).
They process input tensors 
from Python code into kernel arguments, configure launch parameters (\eg, number of thread blocks, 
threads per block, shared-memory size), 
and call the kernels. 
CUDA kernels run on the GPU. They define the computation performed 
by each parallel thread. Under CUDA’s execution model, 
a kernel runs across many threads organized into blocks and grids. 
Each thread uses built-in identifiers (\eg, \texttt{blockIdx.x}, \texttt{threadIdx\allowbreak.x}) 
to determine its location in this hierarchy and 
the portion of input data it processes.

Function \texttt{dynamic\_scaled\_fp8\_quant()} in Figure~\ref{fig:example-c} 
is a wrapper function. It first computes the token count
and feature-element count 
in tensor \texttt{input} on lines 2--3. 
It then sets the number of blocks to the token count on line 4 
and the number of threads per block to 1024 on line 5. 
Finally, it launches kernel function \texttt{scaled\_fp8\_quant\_kernel()} 
with pointers to the underlying memory of the three input tensors and the feature-element count 
on lines 7--8.

Figure~\ref{fig:example-d} shows the CUDA kernel, which 
converts high-precision element values in \texttt{input} into FP8 
using the scaling factor \texttt{scale}. 
To improve efficiency, 
the kernel applies vectorization. On lines 8--9, 
the pointers \texttt{input} and \texttt{out} are cast to vectorized types
so each access processes four elements. Line 10 computes the number of such accesses by dividing 
the element count by four.
The loop on lines 12--16 distributes work across threads 
using a grid-stride pattern. Each thread starts from its global index (\texttt{blockDim.x} * \texttt{blockIdx.x} + \texttt{threadIdx.x}, line 3) and moves 
with a stride equal to the total number of threads  (\texttt{blockDim.x} * \texttt{gridDim.x}, line 5). In each loop iteration, 
a thread loads four input elements on line 13, 
converts them to FP8 using the reciprocal of \texttt{scale}, and writes the results back to the output buffer on line 14.

\subsection{Kernel Memory Bugs in Inference Systems}
\label{sec:safety}

CUDA kernels in LLM inference systems are prone to memory bugs.
Kernels perform index arithmetic over large multidimensional
tensors across thousands of threads;
even small indexing mistakes can cause out-of-bounds accesses or data races.
Performance optimizations (\eg, vectorization, tiling) further complicate this arithmetic.
Moreover, because models and kernels are developed independently
with no formal interface contract,
a kernel that is correct for one model may silently
produce memory errors for another.

\begin{table}[t]
\centering
\footnotesize
\caption{Summary of 20 kernel memory bugs in vLLM.}
\label{tab:vllm-bugs}
\vspace{-0.1in}
{
\begin{tabular}{|c|c|c|}
\hline
\textbf{Dimension} & \textbf{Breakdown} & \textbf{Count} \\
\hline
\hline
\multirow{3}{*}{\textbf{How discovered}}
  & User-reported (production failure) & 13 \\ \cline{2-3}
  & Unit test & 6 \\ \cline{2-3}
  & Code review & 1 \\ \hline
\multirow{2}{*}{\textbf{Memory location}}
  & GPU global memory (tensors) & 17 \\ \cline{2-3}
  & GPU shared memory & 3 \\ \hline
\multirow{3}{*}{\textbf{Bug type}}
  & Integer overflow $\rightarrow$ buffer overflow & 10 \\ \cline{2-3}
  & Buffer overflow (other causes)& 8 \\ \cline{2-3}
  & Null-pointer dereference & 2 \\ \hline
\multirow{2}{*}{\textbf{Patch location}}
  & Kernel code & 18 \\ \cline{2-3}
  & Framework code & 2 \\ \hline
\end{tabular}
}
\end{table}

Figure~\ref{fig:example-d} illustrates a representative bug rooted in the mismatch between model assumptions and kernel interface constraints. 
The kernel uses signed 32-bit integers for \texttt{tid} on line 3 and \texttt{step} on line 5. 
Both are computed as the product of 
the token count (\texttt{blockIdx.x} or \texttt{gridDim.x})
and 1024 (\texttt{blockDim.x}),
implicitly capping the safe token count at 
$2^{21}-1$. 
Neither the model nor the inference framework enforces this bound, as the kernel, model, and framework 
are developed independently with no formal interface constraints on valid input ranges. 
A combination of batch size $BS$ and sequence length $SL$ 
satisfying $BS \times SL > 2^{21}$ (\eg, $BS=116$, $SL=18{,}079$) thus silently triggers overflow, making \texttt{tid} and \texttt{step} negative. 
The corrupted \texttt{tid} is then used as an index to read and write the input/output buffers, 
turning a silent arithmetic error into out-of-bounds memory accesses. 
We reproduced this bug under vLLM on an NVIDIA B300 GPU: 
the overflowed read on line 13 lands on an unmapped virtual address, raising an ``illegal memory access'' fault and crashing the inference service.

To quantify the problem, we examine vLLM's change logs and identify
20 fixed memory bugs in CUDA kernels (Table~\ref{tab:vllm-bugs}).
Four findings stand out.
First, memory bugs in CUDA kernels are common and costly.
Of the 20 bugs, 13 were reported by users after causing failures
in production; only six were caught by unit tests
and one by code review.
Second, nearly all bugs involve tensors.
17 bugs occur on input tensors in GPU global memory
and three in GPU shared memory,
confirming that tensor shape and layout mismatches
are the dominant root cause.
Third, integer overflow is the leading bug pattern.
Ten of the 20 bugs are integer overflows in index computations
that subsequently trigger buffer overflows,
reflecting the growing risk as models and input sequences scale up.
Fourth, fixing these bugs is not always limited to changing the kernel code. Two bugs 
were fixed by modifying the framework's kernel invocation instead. 
This shows that diagnosing and fixing kernel memory bugs requires 
reasoning across the entire software stack, 
rather than analyzing the CUDA kernel alone.
These findings motivate \Tool{}'s design:
\DTool{} makes tensor-level interface constraints explicit,
and \STool{} uses symbolic execution to
check index arithmetic and buffer bounds
that existing testing misses.

\section{Overview of \Tool{}}
\label{sec:overview}

\Tool{} detects memory bugs in CUDA kernels
used by LLM inference systems.
As shown in Figure~\ref{fig:overview}, it consists of two components:
\DTool{}, which constructs constraints that capture
the model-kernel interface,
and \STool{}, which performs symbolic execution on CUDA kernels
under these constraints to identify memory bugs.

\DTool{} supports two inference backends,
vLLM~\cite{vLLM} and Transformers~\cite{transformers},
covering most Hugging Face~\cite{huggingface} models.
Given a model, \DTool{} traces its execution without
launching GPU kernels, and infers whether each kernel argument
is model-fixed or user-controlled and possible relationships among the arguments. 
Then, it synthesizes 
interface constraints
to encode the inference results.
For example, when analyzing the model 
in Figure~\ref{fig:example-a} with the CUDA kernel in Figure~\ref{fig:example-d}, 
\DTool{} infers that tensors \texttt{out} and \texttt{input} in Figure~\ref{fig:example-c}
share the same shape, with their last dimensions 
fixed at 896 (the model's hidden
size) and their first 
dimensions equal to the number of user-input tokens.

\STool{} takes the interface constraints as input and 
checks for out-of-bounds accesses, integer overflows,
data races, and null-pointer dereferences,
reporting each bug's line number along with a triggering input.
For example, 
for the CUDA kernel in Figure~\ref{fig:example-d}
and its wrapper in Figure~\ref{fig:example-c},
\STool{} reports that batch size 116 and
sequence length 18{,}079 
(whose product is the number of input tokens and equals the first dimension of
\texttt{input} and \texttt{out} in Figure~\ref{fig:example-c}) 
trigger integer overflows on lines 3 and
5 of Figure~\ref{fig:example-d} 
and out-of-bounds accesses on lines 13 and 14.

\Tool{} serves two complementary usage scenarios.
Model developers can run it to verify existing kernels
correctly support a new model architecture
and expected token counts,
while kernel developers can confirm that an
optimized or rewritten kernel remains safe across all deployed models. 

\section{\DTool{}: Dynamic Model Profiling}
\label{sec:dynamic}

%
%
As illustrated in Figure~\ref{fig:overview}, \DTool{} operates in four stages. 
It first performs static analysis to identify all CUDA kernels an input model may call. 
It then executes the model using mocked CUDA kernels with a real model configuration, recording how the kernels are actually invoked. 
Next, it mutates the model configuration 
to exercise additional kernels 
identified statically but not triggered dynamically. 
Finally, it aggregates the collected information in a post-mortem analysis to derive constraints that summarize
the interface between the model and each kernel it uses.


\subsection{Static Call Graph Computation}

Given a model's Python implementation 
(\eg, Figure~\ref{fig:example-a}), 
we build a static call graph rooted at the model’s \texttt{forward()} 
method.
The analysis is path- and field-sensitive and leverages 
common coding patterns of Hugging Face models
(\eg, assuming a model object has no aliases 
within its own implementation). 
The analysis over-approximates the results and includes all kernels 
that may be invoked. These kernels guide the subsequent model configuration 
mutation. 
During dynamic profiling, 
\DTool{} ignores any kernels that are not actually executed.
Existing Python call-graph tools cannot correctly 
analyze Hugging Face models~\cite{fraserpyan,multilangdepends,PyCG}.
For example, PyCG does not handle \texttt{\_\_call\_\_}, 
preventing it from resolving calls to an \texttt{nn.Module} object---the standard way to invoke a layer’s \texttt{forward()} 
method (\eg, line 8 in Figure~\ref{fig:example-a}).

Our analysis operates on the Abstract Syntax Tree (AST) of the input Python program.
It maintains an execution context where each program object is represented as an instance, created when its class constructor is analyzed. 
Each instance stores its fields and their values in a dictionary. 
The analysis processes each instruction with respect to this context and also updates
it based on the instruction’s semantics.


Our analysis initializes the context from the model’s \texttt{\_\_init\_\_()} method.
It then analyzes the model's \texttt{forward()} method to construct the call graph.
For each field assignment
(\eg, line 4 in Figure~\ref{fig:example-a}), 
the analysis updates the left-hand instance’s dictionary 
with either the concrete value or the right-hand-side instance.
For each function call, it performs interprocedural analysis by mapping the actual-argument instances to formal parameters.
If the receiver’s type is known only at the base class level, 
all possible subclasses' methods are treated as potential callees.
At each branch, the analysis inspects the context to decide feasible paths.
If both paths may occur, it clones the context and analyzes each path independently.
When encountering Python bindings to CUDA kernels (\eg, line 23 in Figure~\ref{fig:example-a}),
it records that the model calls the corresponding CUDA kernels. 
All Python bindings and their links to CUDA kernels (\eg, Figure~\ref{fig:example-b}) are identified separately using regular expressions.

{
\begin{figure}[t]

\begin{minipage}{\columnwidth}
\begin{center}
\scriptsize

\begin{lstlisting}[style=CodePython,
keepspaces=true,
]
 class Fp8LinearOp:
   def apply(self, input, out_dtype):
     in_2d = input.view(-1, input.shape[-1])
     qinput = torch.empty(in_2d.shape, dtype=out_dtype)
     scale = torch.zeros(1, dtype=torch.float32)
-    torch.ops._C.dynamic_scaled_fp8_quant(qinput, in_2d, scale)
+    stub_dynamic_scaled_fp8_quant(qinput, in_2d, scale)
     return w8a8_scaled_mm_func(...)

+def stub_dynamic_scaled_fp8_quant(output, input, scale):
+  LOG('dynamic_scaled_fp8_quant')
+  LOG(retrieve_stack_info())
+  LOG('output', type(output), output.shape, output.dtype)
+  LOG('input', type(input), input.shape, input.dtype)
+  LOG('scale',type(scale), scale.shape, scale.dtype)
+  return None
\end{lstlisting}
\vspace{-0.1in}
\mycaption{fig:op}{A tracing stub example.}
{}
\end{center}
\end{minipage}
\vspace{-0.3in}
\end{figure}
}



\subsection{Profiling Backends}

We modify the vLLM and Transformers inference frameworks 
to serve as \DTool{} backends and record runtime information during inference.
To speed up profiling, we disable all GPU computations, 
allowing models to run without GPU hardware.
Overall, we make four types of changes.

First, for every registered CUDA kernel, 
we replace its Python binding with a tracing stub. 
As shown in Figure~\ref{fig:op}, the stub function records 
the call stack and parameter types.
For tensor inputs (\eg, \texttt{input}), it
further records their shapes and data types. 
For integer-valued tensors, it also records the minimum and maximum values in the tensors to capture
possible index ranges. 
For primitive-type parameters, the stub records their concrete values. 
If the original binding returns a tensor, 
the stub returns a zero-filled tensor 
with the same shape and data type.

Second, model-loading functions are modified to 
create tensors with correct shapes without loading actual weights.

Third, although no GPU computations are executed, 
we ensure the frameworks follow the same control flow 
as with real GPUs. 
This involves replacing \texttt{vllm.\_C} with a fake module,
since it cannot be imported without a real GPU, and 
adjusting tensors’ internal properties so the frameworks treat  
their data as GPU-resident.

Fourth, we optimize vLLM’s KV-cache management to initialize tensors 
with minimal memory 
and expand them to the required shapes without more memory allocation.
Extra memory is allocated only when a tensor receives writes 
beyond its initial space.
This lazy allocation strategy eliminates pre-allocated memory 
and reduces memory pressure. 

\subsection{Model Configuration Mutation}
For each input model, we mutate two types of execution configuration parameters to trigger CUDA kernels identified statically but not exercised during earlier profiling.

First, several parameters in the model's \emph{config.json} 
govern which CUDA kernels are invoked 
during the forward pass. 
For example,
changing parameter \texttt{quantization\_config} 
selects a different quantization kernel, while reducing \texttt{n\_routed\_experts} from 256 to 128 alters the padding kernel used for token routing from \texttt{sgl\_moe\_align\_block\_size()} 
to \texttt{moe\_align\_block\_size()}.
Because parameter names and semantics vary across models, 
we use an LLM
in agent mode to generate modified config.json files:
given the model’s Hugging Face repository URL and the name of a target kernel, 
the LLM inspects the model code, references the example config.json, 
and produces a new config.json file. 
Not all generated config.json files are valid. 
For example, the model in Figure~\ref{fig:example-a} requires the activation function \texttt{silu}, and changing the \texttt{hidden\_act} field in config.json will cause an assertion error at runtime.
We simply discard any modified config.json 
that results in a runtime failure.

Second, framework environment 
variables and startup parameters also affect 
which kernels are used. 
We maintain a one-time, framework-specific mapping from each CUDA kernel to its required framework configuration.
To trigger a kernel, we retrieve and apply its configuration from this mapping.
If a kernel requires coordinated changes to both framework settings and config.json, 
we apply the framework settings first, then ask the LLM to modify config.json.

\subsection{Interface Constraint Generation}

Given a configuration, \DTool{} executes the model with multiple combinations 
of batch size and sequence length, whose product equals the number 
of input tokens. These two parameters are varied 
because they are controlled by users.
\DTool{} then uses the recorded dynamic information to 
construct the implicit interface
between the model and each called CUDA kernel
as the conjunction of a set of symbolic constraints.
Because the same kernel may be invoked under different call contexts
within a single forward pass,
\DTool{} uses call-stack information
to group invocations and infers constraints
independently for each group.

For each kernel, \DTool{} builds interface constraints over its wrapper
function's arguments, including tensor dimensions, tensor value ranges, and primitive parameters. 
These constraints fall into four categories.

\indentitalicparagraph{Model-fixed constants.}
Some arguments are determined
by the model architecture (\eg, hidden dimension).
\DTool{} identifies them by checking whether a value
is invariant across all runs
and constrains it to the observed constant.

\indentitalicparagraph{User-controlled linear arguments.}
Among the arguments \\that change across runs,
\DTool{} isolates those that scale linearly
with batch size, sequence length, or token count
by verifying that the ratio to the corresponding value
is constant across runs.
These arguments are directly controllable by end users,
and \STool{} treats them as the primary source of
attacker-controlled inputs during verification.

\indentitalicparagraph{Inter-argument equalities.}
Kernel semantics impose dependencies among arguments.
For example, a matrix multiplication requires
the inner dimensions of its operands to match.
\DTool{} detects arguments that always share the same value
and links them with equality constraints.

\indentitalicparagraph{Residual upper bounds.}
Remaining arguments may still depend on the model architecture
indirectly. 
\DTool{} bounds them 
using the maximum value observed across all runs.


For example, after profiling the model in Figure~\ref{fig:example-a}
with a configuration from~\cite{qwen2config},
\DTool{} generates constraints stating that \texttt{out} and \texttt{input}
in Figure~\ref{fig:example-c} share the same shape, that their second dimensions 
are model-fixed at 896 (the hidden size),
that their first dimensions are linear in the token count,
and that \texttt{scale} contains exactly one element.

\section{\STool{}: Symbolic Execution on CUDA}

As shown in Figure~\ref{fig:overview}, 
\STool{} begins its analysis of a CUDA kernel (\eg, Figure~\ref{fig:example-d}) from the kernel's 
wrapper function (\eg, Figure~\ref{fig:example-c}),  
collecting information about 
how the kernel is invoked.  
It then performs symbolic execution on the kernel to detect potential memory bugs.
\STool{} allows users to specify constraints on the wrapper’s inputs, 
so that the interface constraints generated by \DTool{} can be incorporated,
improving both accuracy and efficiency of the symbolic execution. 
Furthermore, \STool{} performs additional validation of detected bugs to ensure they can be triggered under the target inference framework and GPU environment. 
Overall, \STool{} addresses three key challenges.

\begin{table}[!t]
\centering
\footnotesize
\mycaption{tab:tensor}
{Constraint variables defined for tensor \texttt{t}.}{}
\setlength{\tabcolsep}{1.2mm}
\begin{tabular}{|c|c|c|}
\hline
 \textbf{Tensor Fields} & {\textbf{Description}} & {\textbf{Variables}}  \\

\hline
\hline

       \texttt{t.storage\_.data\_ptr\_} &  base address       &   $B_t$  \\ \hline
       \texttt{t.numel\_} &  \# of elements     &   $N_t$ \\ \hline
       \texttt{t.sizes\_and\_strides\_.size\_} &  \# of dimensions &   $D_t$ \\ \hline
       \texttt{t.sizes\_and\_strides\_} &  dimensions    & $d_t^0$, $d_t^1$, ... $d_t^{D-1}$\\ \hline 
    \texttt{t.data\_type\_.itemsize\_} &  element size &   $S_t$  \\ \hline

\end{tabular}

\vspace{-0.1in}
\end{table}

First, prior symbolic execution techniques (\eg, KLEE~\cite{klee}) 
struggle to handle buffers whose sizes are determined at runtime.
In contrast, most CUDA kernels in LLM inference systems operate on the raw memory of tensor objects (\eg, \texttt{out} and \texttt{input} on line 2 of Figure~\ref{fig:example-d}), whose shapes vary depending on model architectures or user-provided tokens.
How can \STool{} analyze tensors with dynamic shapes and memory operations
on the tensors?

Second, wrapper functions often call tensor 
methods to set up parameters (\eg, \texttt{input.size(-1)} on line 2 of Figure~\ref{fig:example-c}) for kernel launches.
These methods often involve complex memory management, and analyzing 
them in full would significantly increase execution time and reduce precision.
How can we summarize these tensor methods to make \STool{} both efficient and accurate?

Third, CUDA kernels run across many threads, 
and analyzing each thread separately would result in redundant work.
How can we share symbolic execution results across threads to avoid unnecessary computations?

\subsection{Memory Model of \STool{}}

\STool{} treats tensor memory separately from other memory regions.

\indentbolditalicparagraphnospace{Tensor Memory.}
All input tensors are stored in GPU global memory
and allocated by the inference systems independently of the kernels.
Therefore, kernel programmers cannot assume any specific memory layout among tensors.
In practice, they never use a pointer to an element of one tensor
to refer to an element of another tensor after arithmetic computations. 
We leverage this property by modeling tensors’ underlying memory regions as discrete 
and far away from each other.

This design contrasts with KLEE's memory model,
which treats the entire heap as a single continuous memory space.
In KLEE, if a buffer has a symbolic size, 
the base addresses of all subsequent buffers become symbolic as well, 
making memory-access reasoning intractable.
This is the fundamental reason why KLEE cannot handle
buffers whose sizes are determined at runtime.

\STool{}'s modeling strategy greatly simplifies 
the detection of out-of-bounds accesses on tensor memory. 
Because each tensor occupies a distinct memory region,
every pointer derived from a tensor carries constraints
that relate it to that tensor's base address.
When a pointer is dereferenced,
\STool{} uses these constraints to identify the source tensor
and check the access against its bounds directly,
without scanning all allocated memory objects---as
a conventional symbolic execution engine (\eg, KLEE) must do
to determine which object a pointer may reference.
For example, when analyzing line 14 in Figure~\ref{fig:example-d}, \STool{}
immediately determines that the memory write must be
checked against the bounds of \texttt{out}.
This follows the constraint established on line 9, which binds \texttt{v\_out} to the base address of \texttt{out}.

\begin{table}[t]
\centering
\footnotesize
\mycaption{tab:methods}
{Tensor-method handling examples.}
{\textit{('-': ignored.)}
}

{
\begin{tabular}{|c|c|c|}
\hline
 \textbf{Category} &  {\textbf{\texttt{Tensor} Methods}} & {\textbf{Constraints}} \\

\hline
\hline

 I & \texttt{t.numel()} & $N_{t}$                   \\ \hline
 I & \texttt{t.data\_ptr()} & $B_{t}$               \\ \hline
  II & \texttt{t2=t1.sum(0)} & \makecell{ $N_{t2}=N_{t1}/d^0_{t1}$,\\ $D_{t2}=D_{t1}-1$, $S_{t2}=S_{t1}$,\\ $d^0_{t2}=d^1_{t1}$, ..., $d^{D-1}_{t2}=d^{D}_{t1}$}               \\ \hline
 III & \makecell{\texttt{check\_dim\_size}\\ \texttt{(t, n, i, exp)}} & \makecell{$D_{t}=n$, $d^i_{t}=exp$}               \\ \hline
  IV & \texttt{t.toString()} & -               \\ \hline

\end{tabular}
}
\vspace{-0.20in}
\end{table}

\indentbolditalicparagraphnospace{Other Memory.}
For global and local memory allocated inside wrapper and kernel functions, 
we use the contiguous memory model adopted by KLEE. 
CUDA shared memory supports both dynamic allocation at kernel launch sites
and static allocation inside kernels, 
and \STool{} handles these two cases separately. 
For dynamically allocated shared memory, \STool{} infers a concrete or symbolic size from the kernel launch parameters. 
For statically allocated shared memory, \STool{} obtains the fixed size directly from the kernel 
declaration. \STool{} models global, shared, and 
local memory as separate regions and 
resolves pointer targets only within the region to which they belong.

\subsection{Handling Tensor Methods}
\label{sec:methods}

To handle tensor methods, 
\STool{} associates each tensor with a set of symbolic constraint variables,
representing its base address, element count, dimensions, and other properties in Table~\ref{tab:tensor}.
Since the number of dimensions
may remain unknown throughout the analysis,
\STool{} maintains a map to record all known dimensions, 
where the keys are dimension indices and the values are the corresponding symbolic variables (\eg, $d_{1}$, $d_{2}$). 
For example, \texttt{input.size(-1)} on line 2 of Figure~\ref{fig:example-c} accesses the last dimension of tensor \texttt{input}.
\STool{} processes this method by adding an entry to the map with key $-1$  and value $d^{D-1}_{input}$. 

With the constraint variables, \STool{} supports 140 \texttt{Tensor} and \texttt{TensorIterator} methods, 
grouped into four categories. 
Table~\ref{tab:methods} shows examples. 
\textcircled{1}~\emph{Property queries} (36 methods)
return tensor metadata such as element count or shape.
\STool{} models their return values
using the symbolic variables in Table~\ref{tab:tensor}.
For example, \texttt{input.numel()} on line 3 of Figure~\ref{fig:example-c}
returns 
$N_{input}$.
\textcircled{2}~\emph{Tensor-producing methods} (70 methods)
create a new tensor or a sub-tensor.
\STool{} introduces a set of fresh symbolic variables 
for the result
and adds constraints relating it to the original.
For example, \texttt{t2 = t1.sum(0)} reduces \texttt{t1} along its first dimension,
so \texttt{t2} has one fewer dimension and its $i$-th dimension
equals the $(i{+}1)$-th dimension of \texttt{t1} (Table~\ref{tab:methods}, Row 3).
\textcircled{3}~\emph{Property checks} (4 methods)
inspect tensor properties (\eg, contiguity);
\STool{} encodes the success condition as a constraint.
\textcircled{4}~\emph{Analysis-irrelevant methods} (30 methods)
do not affect bug detection and are skipped.

\begin{table}[t]
\centering
\footnotesize
\mycaption{tab:thread}
{Constraint variables defined for CUDA threads.}{}
\setlength{\tabcolsep}{1.2mm}
\begin{tabular}{|c|c|c|}
\hline
 \textbf{Variables} & {\textbf{Description}} & {\textbf{Type}}  \\

\hline
\hline

        $blockIdx.x/y/z$ &  block identifier     &   ID  \\ \hline
        $threadIdx.x/y/z$ &  thread identifier     & ID   \\ \hline
        $gridDim.x/y/z$ &  grid size     &   launch \\ \hline
       $blockDim.x/y/z$ &   block size    &  launch \\ \hline

\end{tabular}

\vspace{-0.15in}

\end{table}

\subsection{Handling CUDA Threads}
\label{sec:threads}

CUDA kernels execute across thousands of threads
organized into blocks and grids.
All threads run the same code but diverge based on
their thread and block identifiers,
which programmers also use to partition work over input data.

Rather than enumerating threads,
\STool{} models thread and block identifiers as symbolic variables,
so that a single symbolic execution pass captures the behavior of all threads.
As shown in Table~\ref{tab:thread},
\STool{} introduces six ID variables
(\eg, \texttt{threadIdx.x}, \texttt{blockIdx.y})
and six launch variables
(\eg, \texttt{blockDim.x}, \texttt{gridDim.y}).
ID variables are bounded by launch variables
following CUDA semantics (\eg, $blockIdx\allowbreak.x < gridDim.x$),
and launch variables are concretized or constrained
when \STool{} analyzes kernel launch sites
(\eg, $gridDim\allowbreak.x = num\_tokens$
in Figure~\ref{fig:example-c}).
With this approach,
thread-independent bugs are detected as in sequential programs;
thread-dependent bugs (\eg, line 13 in Figure~\ref{fig:example-d})
are captured through constraints over the ID variables.

For loops that distribute work across threads 
(\eg, the highlighted loop in Figure~\ref{fig:example-d}),
\STool{} analyzes their loop bodies once
and then adds the loops’ exit conditions to its execution states
(\eg, $i \geq num\_vec\_elems$)
for the remaining execution.
This suffices because such loops follow a simple strided pattern: each thread starts at an offset given by its global thread index 
and advances by the thread count. 
A single symbolic analysis of the loop body can capture every loop index
a thread may process, while the exit condition characterizes the post-loop state.



To detect data races, \STool{} introduces a second set of
six ID variables representing a distinct thread.
It asserts that the two threads differ
by negating the conjunction of pairwise equality
over their ID variables.
For each global-memory access,
\STool{} checks whether the second thread
may perform a conflicting write to the same address.
\STool{} performs a similar check for shared memory but additionally constrains the two threads' block ID variables to be equal, 
since shared memory is private to each block.

\subsection{Synthesizing Input and Buggy Constraints}
\label{sec:constraints}

\indentbolditalicparagraphnospace{Input Constraints.}
Before analyzing each wrapper function,
\STool{} introduces three symbolic variables:
$BS$ (batch size), $SL$ (sequence length),
and $TC$ (token count), related by $TC = BS \times SL$.
In addition to the interface constraints from \DTool{},
\STool{} bounds these variables using two extra sources:
\textcircled{1}~\emph{Model-defined limits.}
Many models specify a maximum sequence length
through fields such as ``\texttt{max\_position\_embeddings}'',
``\texttt{n\_positions}'', or ``\texttt{max\_seq\_len}'' 
in their config.json files.
\DTool{} extracts these values from each input 
model's config.json
and \STool{} uses them to upper-bound $SL$.
\textcircled{2}~\emph{Practical input limits.}
Even when a model does not specify a maximum,
the sequence length is bounded in practice.
Current production models (\eg, GPT-5.4~\cite{gpt5.4}, Claude Opus 4.6~\cite{claude4.6}, Gemini 3.1~\cite{gemini3.1}) support up to roughly one million tokens.
We therefore impose default bounds of $BS \leq 1{,}000$
and $SL \leq 1{,}000{,}000$,
which can be adjusted for specific deployment scenarios.


\indentbolditalicparagraphnospace{Buggy Constraints.}
\STool{} checks four types of memory bugs
(data-race detection is described in Section~\ref{sec:threads}).
For \emph{out-of-bounds accesses},
\STool{} adds a constraint at each memory access via a pointer
asserting that the pointer falls outside
the valid range 
of the accessed buffer.
For \emph{integer overflows},
\STool{} adds a constraint at each 32-bit addition or multiplication
asserting that the result exceeds the type's maximum value.
For \emph{null-pointer dereferences},
\STool{} adds constraints for each pointer dereference indicating 
the pointer may have been computed from an unknown base pointer. 
All input tensors are assumed to be valid
with base addresses properly initialized by the inference
frameworks.


\subsection{Bug Detection and Validation}

\STool{} begins symbolic execution from each wrapper function under the
input and interface constraints, processing tensor methods
(Section~\ref{sec:methods}) and, upon each kernel launch, introducing thread
ID variables and launch variables constrained by CUDA semantics
(Section~\ref{sec:threads}).
\STool{} then executes the kernel body instruction by instruction. 
At each potentially buggy instruction, it 
synthesizes buggy constraints
(Section~\ref{sec:constraints}), conjoins them with existing constraints,
and invokes a solver.
A satisfiable result yields concrete values for all symbolic variables,
including batch size $BS$ and sequence length $SL$.

However, not every satisfiable assignment is triggerable in practice.
$BS$ and $SL$ are end-user inputs to the inference framework, 
but the framework may reject certain values, 
or the resulting memory footprint may exceed GPU capacity. 
\STool{} checks both conditions to confirm a bug.

We replay the $(BS, SL)$ pair via \DTool{}'s profiling backend.
If the kernel is still invoked
and its arguments match the interface constraints,
the bug is confirmed to be reachable through the framework.

\begin{table}[!t]
\centering
\footnotesize

\mycaption{tab:detection}
{Benchmark information and detection results.}
{\textit{(
IO: integer overflow,
OOB: out-of-bounds,
DR: data race,
and NULL: null-pointer dereference.
$x_{y}$: x true positives, y false positives.
'-' indicates zero true and false positives.)}
}
{
\setlength{\tabcolsep}{1.2mm}
\begin{tabular}{|l|c|c||c|c|c|c|c|}
\hline
    & {\textbf{Models}} & {\textbf{Kernels}} & {\textbf{IO}} & {\textbf{OOB}} & {\textbf{DR}} & {\textbf{NULL}} & {\textbf{Total}}   \\ 

\hline
\hline

{\textbf{vLLM}}  & $62$ & $50$   & $7_{2}$     & $19_{5}$  & $0_{2}$ & - & $26_{9}$  \\ \hline
{\textbf{HF}}  & $54$ & $117$ & $ 131_{0}$     &  -         & - & - & $ 131_{0}$  \\ \hline
{\textbf{Papers}}  & $4$ & $6$ & $24_{0}$      &  -         & - & - & $24_{0}$    \\ \hline  \hline
{\textbf{Total}}  & $120$  & $173$ & $162_{2}$  & $19_{5}$  & $0_{2}$ & - & $181_{9}$ \\ \hline

\end{tabular}
}

\vspace{-0.15in}
\end{table}

We estimate GPU memory consumption as model-weight storage
plus KV-cache size, scaled by a $1.25\times$ fragmentation factor for
the vLLM backend and $1.45\times$ for Transformers.
Model-weight size is obtained directly from the weight files.
KV-cache size is
$2 \times L \times TC \times H \times \mathit{sizeof}(\mathit{dtype})$, 
where $L$ is the number of layers, $TC$ is the token count,
$H$ is the hidden dimension, and $\mathit{dtype}$ is the cache data type.
Intermediate tensors and temporary buffers are negligible
relative to weights and KV-cache and are omitted.
If the estimated total exceeds the target GPU's memory capacity,
the bug is marked as untriggerable.
For the bug in Figure~\ref{fig:example-d},
the solver reports $BS{=}116$ and $SL{=}18{,}079$;
the estimated memory usage is 211.2GB,
well within a B300 GPU's 288GB memory capacity.
We confirm the bug is triggerable on this hardware.

\section{Evaluation}
\label{sec:eval}

\indentbolditalicparagraphnospace{Implementation.}
\DTool{} is implemented in Python. 
It performs call-graph analysis using the built-in AST module~\cite{ast-python},
provides two profiling backends (vLLM-0.9.0~\cite{vLLM} 
and Transformers v4.52.1~\cite{transformers}), 
and employs ChatGPT agent~\cite{gpt-agent} to mutate config.json files.
The agent has web search capability to retrieve relevant code.
In total, \DTool{} comprises 18,453 lines of code, including both standalone modules 
and framework modifications. 
\STool{} is built on KLEE v3.1 and operates on LLVM IR generated from CUDA source code.
We implement 19,572 lines of C++ to support 
tensor abstractions and enable symbolic execution on CUDA threads and 
GPU instructions.
\STool{} uses Z3~\cite{Z3} as its constraint solver.


\indentbolditalicparagraphnospace{Research Questions.}
Our experiments aim to evaluate \Tool{}'s effectiveness 
in detecting memory bugs in CUDA kernels used by real-world inference systems 
(Section~\ref{sec:effectiveness}), its coverage of CUDA memory bugs (Section~\ref{sec:coverage}), its advantages over existing 
techniques (Section~\ref{sec:coverage}), and the rationale behind 
combining dynamic model profiling with CUDA-specific symbolic execution (Section~\ref{sec:rational}).

\indentbolditalicparagraphnospace{Experimental Settings.}
Most experiments run on a server with an Intel Xeon Silver 4110 CPU, 256GB RAM, and Red Hat Enterprise Linux 9. 
We compile CUDA code for CUDA 12.1 using clang++ 13.0.0 with ``\texttt{-emit-llvm}'' 
to generate LLVM IR for \STool{}. Unless stated otherwise, 
we target an NVIDIA B300 GPU with 288GB memory for bug validation.

\subsection{Bug Detection in the Wild}
\label{sec:effectiveness}

\subsubsection{Methodology}
\label{sec:effective_meth}

As shown in Table~\ref{tab:detection}, we evaluate \Tool{} 
on models and kernels from three real-world sources.
First, we extract all text-generation models and their associated kernels from vLLM-0.9.0, 
the latest release at the start of this project, 
yielding 62 models and 50 kernels, with substantial kernel reuse across models. 
Because vLLM omits \textit{config.json} files, we retrieve them from Hugging Face by querying model IDs specified in vLLM’s tests,  
and selecting the config.json from the top-ranked repository for each model. 
Second, we gather all Hugging Face models that contain custom CUDA kernels, resulting in 54 models and 117 unique kernels.
Third, we survey top-tier AI conference papers from the past two years and identify four~\cite{you2024shiftaddllm,malinovskii2024pvtuning,park2024anyprecision,huang2025mixturecompressor} 
whose artifacts provide both Hugging Face models and custom kernels, contributing four models and six kernels.
In total, our evaluation includes 120 models and 173 
kernels, comprising 170,875 lines of Python model code 
and 139,951 lines of CUDA code.

We evaluate benchmarks from the first source 
using the vLLM backend and the remaining benchmarks using the Transformers backend. For each model, 
we attempt configuration mutation only once for any CUDA kernel that 
is statically reachable but not triggered by the collected config.json. 
For every configuration (both original and mutated), 
we test nine combinations of batch sizes $\{1, 3, 5\}$ 
and sequence lengths $\{1, 7, 17\}$, 
chosen to keep profiling short while remaining realistic and uncommon. 
Each CUDA kernel is analyzed by \STool{} for up to one hour, following common practice in prior symbolic-execution work~\cite{klee, kim2012industrial, li2013steering}.

For each bug reported by \STool{}, 
we examine the corresponding CUDA code to verify that the reported tensor shapes and input values indeed trigger the issue and are semantically valid.
To ensure objectivity, each bug is independently reviewed by at least two authors. 
Because vLLM kernels are shared across models, 
we count unique bugs by kernel line number, even if they are triggered by different models.
For out-of-bounds accesses resulting from a preceding integer overflow, 
we count only the integer overflow.

\subsubsection{Experimental Results}

As shown in Table~\ref{tab:detection}, 
\Tool{} detects 181 memory bugs in CUDA kernels, including 162 integer 
overflows and 19 out-of-bounds accesses. 
It also identifies one benign race in vLLM, which we conservatively classify 
as a false positive. 
\Tool{} does not report any null-pointer dereferences. The large number of bugs demonstrates 
\emph{\Tool{}’s effectiveness in detecting memory bugs in CUDA kernels.}

In addition, \Tool{} detects bugs across benchmarks from all three sources, 
underscoring \emph{its ability to analyze diverse models and kernels}.

We have reported all 26 bugs detected from vLLM, which has an actively maintained GitHub repository. 
Of them, eight have been fixed in the most recent version, and four extra bugs have been confirmed by the developers.
Among the 18 unresolved bugs, the buggy code has been refactored for 13, while the other five are still in the codebase. 
We do not report bugs in Hugging Face models or research papers when the corresponding GitHub repository is unavailable, 
has been inactive for more than a year, or no contact email is found. 
For the 36 remaining cases, we have contacted the authors by email 
but haven’t received any response.

{
\begin{figure}[t]

\begin{minipage}{\columnwidth}
\begin{center}
\scriptsize

\begin{lstlisting}[
        style=CodeCuda,
        escapechar=@,
    ]
// Qwen-1_8B-Chat: heads=16, height=128, hidden_size=heads*height
// batch: batch_size, width: seq_len
__global__ void VecQuant8BatchMatMulKernel_faster_old(const half* vec, const uint8_t* mat, half* mul, int batch, int width, int heads, int height) {
  int weight_total = batch * width * heads * height; // @{\color{red}overflow}@
  int h = 128 * blockIdx.x;
  int w = 128 * blockIdx.y + threadIdx.x;
  int i = width * h + w;
  half2 weight[64];
  for (int b = 0; b < batch; ++b) {
    for (int head = 0; head < heads; ++head) {
      int shift = b * heads + head;
      for (int k = 0; k < 64; ++k) {
        int index1 = shift * height * width + i + (2 * k * width);
        if (index1 >= weight_total || w >= width || ...)
          weight[k] = __float2half2_rn(0);
        else
          weight[k] = compute_weight(mat, ...);
  }}}
  compute_mul(mul, vec, weight,...);
}
\end{lstlisting}
    \vspace{-0.1in}
\mycaption{fig:io-example-hidden}{A detected integer overflow bug.}
{}
\end{center}
\end{minipage}
\vspace{-0.25in}
\end{figure}
}

\indentitalicparagraph{Integer Overflows.}
Three detected integer overflows occur on unsigned 32-bit integers,
and the rest occur on signed \texttt{int32} values. 
In 15 cases, the overflowed value is later used as an index, triggering subsequent 
out-of-bounds accesses.
Figure~\ref{fig:example-d} shows one such example. 

In terms of what is being computed, 
36 of the overflows come from multiplying the token count by the hidden size.
For example, the kernel \texttt{VecQuant8BatchMatMulKernel\_faster\_old()} in Figure~\ref{fig:io-example-hidden} 
multiplies input matrices \texttt{vec} and \texttt{mat}, 
and stores the result in \texttt{mul}.
Since \texttt{mat} is quantized, 
the kernel dequantizes it on lines 9--18 before multiplying on line 19. 
During dequantization, \texttt{weight\_total} is used on line 14 
to determine whether a value should be zeroed.
However, the computation of \texttt{weight\_total} can overflow. 
Specifically, \texttt{batch * width} equals the token count, and \texttt{heads * height} equals the hidden size. While the hidden size is fixed at $16 * 128$, 
overflow can occur when the batch size is 128 and the sequence length is 8,192.
When this happens, \texttt{weight\_total} becomes negative, 
zeroing all dequantized values on line 15 and corrupting the subsequent computation.
Another 98 integer overflows occur when 
computing the element count in $QK^\top$, 
whose size is $\texttt{batch\_size} \times \texttt{num\_heads} \times \texttt{seq\_len}^2$ and 
can exceed the \texttt{int32} limit. 
Four additional 
overflows arise when multiplying the token count by 
1,024 (thread count), as shown in Figure~\ref{fig:example-d}. 
The remaining 24 overflows come from a single code fragment 
that appears 24 times at different stages of a long computation.
In each
fragment, a 32-bit integer is shifted left by 10 bits and 
then cast to a 16-bit integer. When the original value exceeds 31, an overflow occurs during the cast. 
Neither the model code nor the 
CUDA kernel imposes any constraint to prevent this value 
from exceeding 
that limit.
We count these bugs as separate bugs, according to the methodology described in Section~\ref{sec:effective_meth}. 

The proportion of integer overflows among the detected bugs is markedly
higher than in the bugs collected from vLLM’s change log in Section~\ref{sec:safety}. 
We hypothesize that integer overflows 
are often under-reported because overflowed values are typically 
used in computations rather than as indices, 
causing silent result corruption instead of crashes.

\begin{figure}[t] 
    \centering
    \includegraphics[width=0.9\columnwidth]{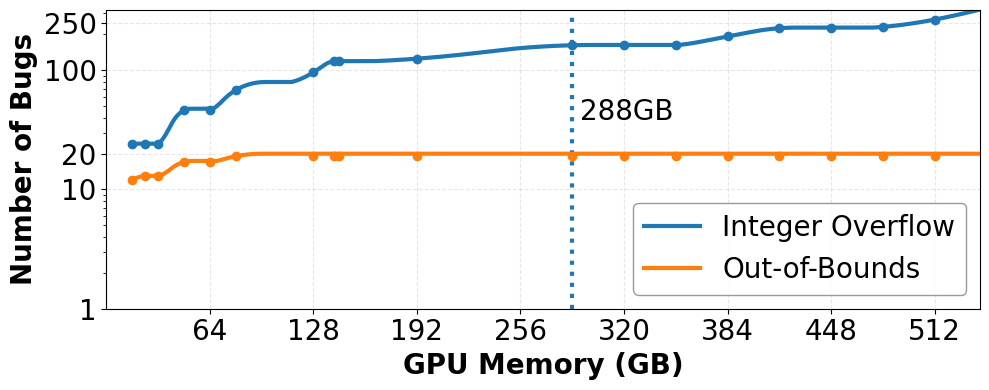} 
\vspace{-0.1in}
\mycaption{fig:mem-bug}{How the number of bugs changes with
GPU memory.}
{\textit{(The vertical line represents NVIDIA B300.)}}
\vspace{-0.1in}
\end{figure}

\indentitalicparagraph{Buffer Overflows.}
Eight buffer overflows occur when data is partitioned into blocks, but the final block is only partially filled, yet kernels access it as if fully populated. 
The remaining 11 arise when input tensor values are used to compute indices into another tensor 
without adequate bounds checking.
For them, we further confirm that the framework code permits the 
buggy tensor content to pass into the kernel. 

\indentitalicparagraph{Data Races.}
\Tool{} detects a data race, where 
an array is used to count tokens assigned 
to each expert, with one array element per expert. 
The array elements are partitioned among warps.
During initialization, all 32 threads in a warp write zero to all the elements in the warp's partition, creating a race. 
We consider the race benign and classify it as a false positive, 
as all conflicting threads write the same value to the same location. 
Nevertheless, this case demonstrates \Tool{}'s capability to detect data races in CUDA kernels.

\indentitalicparagraph{Security Impacts.}
As an initial exploration, we reproduce a subset of the detected bugs 
and examine their potential security impacts. 
A systematic study is left for future work.
Most reproduced integer overflows corrupt computation results and may affect downstream tasks. 
Most reproduced out-of-bounds accesses (\eg, Figure~\ref{fig:example-d}) 
trigger crashes 
because they access unmapped memory.
We also find one case in which a signed-index overflow is used 
in a memory write, modifying a neighboring tensor 
and corrupting the inference results of other users in the same batch. 
We record a video demonstrating this effect~\cite{TokenSmash}. 
In a newer version of vLLM, we find a similar bug in which an overflowed 
unsigned index is used for a memory read, 
exposing another user's inference output. 
This bug was introduced after vLLM-0.9.0 and is therefore not included in Table~\ref{tab:detection}. 
Our finding leads to a published security advisory~\cite{security-advisory},
and we also record a demonstration video~\cite{TokenBleed}.
Finally, we find it difficult to exploit the detected out-of-bounds accesses for arbitrary code execution, since return addresses are 
typically stored in registers during CUDA 
function calls and are not easily overwritten unless register spilling occurs.

\indentitalicparagraph{False Positives.}
\Tool{} reports nine false positives (false discovery rate of 4.74\%), 
arising from three sources.
First, six stem from \DTool{}’s inability to track tensor values across kernels: it returns zero-value tensors
for kernel binding functions, so subsequent kernel bindings cannot track valid value ranges for those tensors, 
causing two integer-overflows, three out-of-bounds, and one data-race false positives.
Second, two involve kernels that index tensors using values loaded from model weights. As \DTool{} does not load the actual weights, it cannot accurately capture the indices; manual inspection confirms 
that the true weight values cannot trigger errors. 
Third, \Tool{} identifies one benign race, as discussed earlier.

\indentitalicparagraph{Memory Bugs vs.\ GPU Memory.} 
Figure~\ref{fig:mem-bug} shows how detected bugs vary with GPU memory size, 
which \STool{} uses to decide whether a bug can be triggered. 
Out-of-bounds bugs plateau beyond 80GB, 
whereas integer overflows grow with larger memory. 
On the NVIDIA H100 (80GB), \Tool{} finds 68 integer overflows 
and 19 out-of-bounds bugs; on the H200 (141GB), integer overflows rise to 119 
while out-of-bounds bugs remain unchanged; 
on the B300 (288GB), integer overflows further increase to 162. 
This trend suggests that future GPUs with larger memory 
will expose more integer overflow bugs.

\indentitalicparagraph{Time and Monetary Cost.}
\Tool{} takes 1,073.62 total hours to complete its analysis: 
1,032.99 for vLLM, 28.18 for Hugging Face, 
and 12.45 for research papers. 
Experiments run with multiple parallel processes, 
and the reported time reflects their aggregate. 
Using GPT costs \$442.16. 
Considering that both the Python and CUDA codebases exceed 100,000 lines, 
\Tool{} incurs moderate time and monetary costs.

More specifically, \STool{} consumes 979.86 hours, 
substantially more than \DTool{} (93.76 hours).
\DTool{} generates 1,274 distinct model configurations. 
Because many vLLM kernels are shared across models, 
we merge identical analysis contexts, yielding 8,562 unique contexts for \STool{}. 
Kernel analysis completes within one hour for 7,784 cases 
and reaches the limit for 778 cases. All kernels exceeding the one-hour limit 
contain 600+ lines of CUDA code.

\boldunderparagraph{Answer to effectiveness:}
{\it{ \Tool{} can accurately detect integer overflows and buffer overflows in CUDA kernels used within LLM inference systems.
}}

\subsection{Coverage and Advancement}
\label{sec:coverage}

\begin{table}[t]
\centering
\footnotesize

\mycaption{tab:compare}
{Comparison with baseline techniques.}
{\textit{(20 bugs are for each row. `/': not applicable.)}
}
{

\setlength{\tabcolsep}{1mm}
\begin{tabular}{|l|c|c|c||c|}
\hline
    & {\textbf{Honeycomb}} & {\textbf{GKLEE}} & {\textbf{ESBMC-GPU}}  & {\textbf{\STool{}}}  \\ 

\hline
\hline

{\textbf{vLLM}}      & / & / & /   & 15     \\ \hline
{\textbf{Simplified}}  & 0 & 5  & 3 &   19    \\ \hline

\end{tabular}
}

\vspace{-0.15in}
\end{table}

\subsubsection{Methodology}
We evaluate coverage using the 20 known vLLM bugs 
from Section~\ref{sec:safety}. Since most of the bugs 
were reported by users without model information, we evaluate only \STool{}.
We retain inputs required by the model architecture, if any, and treat the rest as symbolic
when running \STool{}. 


We compare \STool{} with three static CUDA analysis \\tools: Honeycomb's validator~\cite{mai2023honeycomb}, 
GKLEE~\cite{li2012gklee}, and 
ESBMC-GPU~\cite{pereira2016verifying}. 
These tools cannot run directly on vLLM kernels due to 
incompatibilities with modern C++, CUDA, and PyTorch/vLLM libraries. 
Moreover, GKLEE and ESBMC-GPU require a \texttt{main()} function as their entry point, 
which vLLM kernels lack.
To enable comparison, we simplify the vLLM kernels 
by rewriting modern CUDA and C++ features 
with legacy constructs, 
removing PyTorch and vLLM dependencies, 
and adding a \texttt{main()} function for each kernel that replaces the original wrapper and allocates GPU memory for tensors. 
The simplified kernels preserve the root causes of the bugs.
Because Honeycomb targets HIP kernels~\cite{amdhip}, we additionally translate CUDA APIs to HIP equivalents using 
hipify-perl~\cite{amdhipifyperl}.
We do not compare with other related tools 
because they are not free~\cite{axivioncuda, parasoftcuda} 
or their code is not available~\cite{chatterjee2025proofwright, li2010symbolic, li2014practical, li2013se}.

\subsubsection{Experimental Results}

As shown in Table \ref{tab:compare}, \STool{} identifies 15 of the 20 known bugs, demonstrating 
\textit{strong coverage of real memory errors in CUDA kernels}. 
Specifically, the detected bugs comprise nine integer overflows, five out-of-bounds accesses, 
and one null-pointer dereference, underscoring \textit{\STool{}'s ability to detect the three bug classes}.

Figure~\ref{fig:oob-example} shows an out-of-bounds error. 
The wrapper function (omitted for brevity) allocates shared memory to record how threads assign tokens to experts. As shown on line 2, 
the first \texttt{num\_experts + 1} entries store the total token count and 
the per-expert counts. The remaining entries 
form a 2D array storing
per-thread, per-expert token routing
plus one slot 
per expert for the expert’s total. Line 5 attempts to 
skip this first region so that lines 7--9 operate on the second, 
but incorrectly uses the thread count (\texttt{blockDim.x}),
instead of expert count, to compute the offset. 
As marked on line 1, 
when the expert count is below 32, the thread count 
exceeds it, leading to a buffer overflow on line 8.
\STool{} correctly reports this, 
with \texttt{threadIdx.x=31} and \texttt{num\_experts=8}. 

{
\begin{figure}[t]

\begin{minipage}{\columnwidth}
\begin{center}
\scriptsize

\begin{lstlisting}[
        style=CodeCuda,
        escapechar=|,
    ]
// num_thread = std::max(num_experts, 32)
// share_mem size: num_experts+1+(num_thread+1)*num_experts
__global__ void moe_align_block_size_kernel(int32_t num_experts){
  extern __shared__ int32_t shared_mem[];
- int32_t* tokens_cnts = shared_mem + blockDim.x + 1;
+ int32_t* tokens_cnts = shared_mem + num_experts + 1;
  for (int i = 0; i < num_experts; ++i) {
    tokens_cnts[num_experts * (threadIdx.x + 1) + i] = 0; // |{\color{red}OOB}|
  }}
\end{lstlisting}
    \vspace{-0.1in}
\mycaption{fig:oob-example}{An out-of-bounds bug from vLLM.}
{}
\end{center}
\end{minipage}
\vspace{-0.2in}
\end{figure}
}

The missed bugs arise for four reasons. 
First, 
one null-pointer dereference occurs because 
the kernel accesses data on a different GPU, 
undetectable via static analysis. 
Second, one out-of-bounds is caused by a mismatch between 
a tensor’s declared shape and the underlying memory layout; 
\STool{} assumes consistency between them and 
thus misses the issue. Third, two bugs exceed the one-hour analysis limit 
and remain undetected. 
Fourth, \STool{} lacks support for C++ \texttt{std::map}, 
causing it to miss one bug that depends on this construct. 
For the simplified kernels, \STool{} still misses the 
first bug but successfully identifies the other four.

\indentitalicparagraph{Honeycomb} detects no bugs because 
it only checks whether memory accesses stay within one 
of four coarse memory regions (protected, read-only, read-write, and private). 
All tested out-of-bounds accesses fall inside the read-write region, so Honeycomb cannot catch them.

\indentitalicparagraph{GKLEE} detects four out-of-bounds accesses and one integer overflow, but misses the rest for three reasons. 
First, seven bugs require 
\texttt{threadIdx.x} or \texttt{blockIdx.x} to exceed 30, 
but GKLEE assumes small constant values for both. 
Second, GKLEE fully unrolls loops, causing the analysis of seven bugs to hit the one-hour timeout.
Third, GKLEE cannot detect the bug due to the kernel and its data on different GPUs.

\indentitalicparagraph{ESBMC-GPU} identifies two out-of-bounds accesses and one integer overflow. It misses nine, seven, and one bugs respectively for the same reasons as GKLEE. 

\boldunderparagraph{Answer to coverage and advancement:}
{\it{
\STool{} detects most of the tested known bugs, substantially more than the baselines, demonstrating its strong coverage of real CUDA memory bugs and its advantages over existing techniques.
}}
\subsection{Rationality of Components}
\label{sec:rational}

\subsubsection{Methodology}

We conduct an ablation study by disabling individual components of \Tool{} and comparing each variant against the full system. 
We use the same vLLM models, kernels, and experimental setup as in Section~\ref{sec:effectiveness}, 
making the results directly comparable to those reported there. 
The variants are:
\textcircled{1} \textit{Without \DTool{}.}
We run \STool{} directly on kernel wrapper functions without guidance from \DTool{}.
\textcircled{2} \textit{Without \STool{}.}
We execute each model on real GPUs using both original and mutated configurations, 
and record runtime errors (\eg, ``illegal memory access’’). 
Experiments are parallelized across four RTX 5090s and four H100s; since only error occurrence is checked, hardware differences do not affect results.
\textcircled{3} \textit{Without configuration mutation.} 
We run \Tool{} on the collected config.json files and frameworks' default configurations, 
without any mutations. 
\textcircled{4} \textit{Without validation.}
All bugs reported by symbolic execution are treated as final outputs without validation.

\begin{figure}[!t] 
    \centering
    \includegraphics[width=0.9\columnwidth]{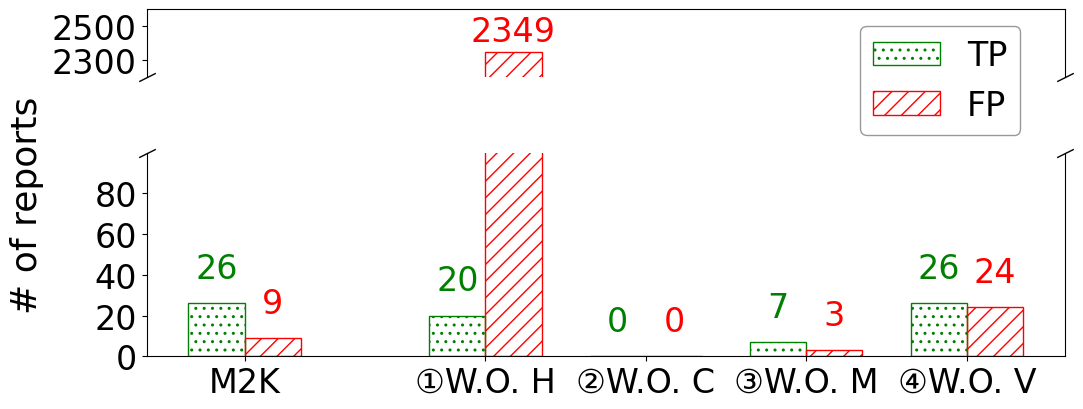} 
\vspace{-0.05in}
\mycaption{fig:ablation}{Contributions of \Tool{}'s components.}
{\textit{(26 bugs detected in vLLM. W.O.: without, C: \STool{}, 
H: \DTool{},
M: configuration mutation, and V: Validation.)}}
\vspace{-0.1in}
\end{figure}

\subsubsection{Experimental Results}

The full-featured version of \Tool{} detects the most 
bugs in Figure~\ref{fig:ablation}, demonstrating \textit{the value of each component}.

Removing \DTool{} causes \STool{} to report 2,349 false positives for two reasons.
First, \STool{} may infer tensor shapes that are theoretically possible but never occur in reality, leading to incorrect out-of-bounds reports.
For example, in Figure~\ref{fig:example-c}, 
\STool{} incorrectly assigns tensor \texttt{scale} a zero size to trigger out-of-bounds access on line 4 of Figure~\ref{fig:example-d}. However, this tensor always has one element in reality, 
making such a case infeasible.
Second, \STool{} may treat tensor shape values or other parameters as having extremely large, unrealistic values (\eg, a hidden size of 2,147,483,640).
Moreover, without model-architecture information, 
all tensor shapes become symbolic, forcing \STool{} to explore far
more possibilities.
As a result, it fails to complete the analysis of 25 
kernels within one hour, 18 
more than when architecture information is available, 
and increases the average analysis time per kernel by $2.1\times$.
The six missed bugs are due to this reason. 

%


With \STool{} disabled, 
no runtime errors are detected, so this variant reports 
nothing. Our experiments use a maximum batch size of 5 
and a sequence length of 17 (only 85 tokens in total), 
far too few to expose hidden bugs. This shows that \Tool{} uncovers issues that would otherwise appear only at large production scales.


Without configuration mutation, only 21 of the 50 CUDA kernels are exercised, leading \Tool{} to miss 19 bugs.


Disabling validation causes
\Tool{} to report 15 more false positives. Five are cases where the token count is too large, causing memory usage to exceed the B300's memory size. 
The remaining ten cases should be filtered out by rerunning the profiling backend with the batch size and sequence length reported by the solver.

\boldunderparagraph{Answer to rationality:}
{\it{
Each component of \Tool{} plays a distinct and essential role in its overall effectiveness.
}}
\section{Limitations and Discussion}
\label{sec:diss}

\Tool{} currently targets Hugging Face models and CUDA kernels with tensor inputs, 
which together cover a large share of real-world usage. 
It profiles kernel usage without executing kernels on any specific GPU, making its effectiveness independent of any GPU hardware. 
\Tool{} also supports GPU-dependent conditional compilation through cross-compilation and can handle fused kernels when the fused source code is available.
At present, \Tool{} does not support Triton kernels, NPU kernels, 
or TensorFlow models. However, applying its core idea of an explicit model-kernel interface to these settings is promising 
for uncovering potential issues.

\Tool{} mutates execution configurations to trigger additional CUDA
kernels. It does not aim to cover all possible
models, nor is this necessary. In practice, users care about the
specific model they deploy, rather than whether a kernel behaves correctly under every possible model.
\Tool{} therefore focuses on the kernel
shapes exercised by a given model. 
Although our results show that the mutation is
effective, it is far from a systematic inspection. 
Developing a more comprehensive inspection method remains future work.

In our experiments, \STool{} analyzes each kernel for up to one hour. 
For certain kernels (\eg, \texttt{gptq\_marlin\_gemm()}), 
not all paths are explored, potentially missing 
memory bugs. 
Since \STool{} inherits KLEE's path exploration 
strategies designed for C/C++, optimizing them for CUDA code patterns 
is key to improving effectiveness.

Most bugs detected by \Tool{} are integer overflows. 
Although such bugs look simple, we are not aware of a simpler way to resolve them. First, a static technique must track the 
numerical relations among variables, 
the path constraints under which these relations hold, 
and the constraints imposed by the model and the framework; without this information, it inevitably reports many false positives. 
Second, a dynamic technique must synthesize kernel inputs satisfying 
the model's constraints and run them on real GPU hardware, 
both of which are hard to automate. Moreover, as our experiments show, most overflows silently corrupt computation results, making them difficult to detect dynamically.
Third, simply using 64-bit integers everywhere is not an ideal solution because it increases both runtime and memory overhead.
\Tool{} overcomes these difficulties by combining dynamic profiling 
with static symbolic execution, enabling it to detect integer overflows both effectively and precisely.

\section{Related Work}
\label{sec:related}


\indentbolditalicparagraphnospace{Exploiting CUDA Memory Bugs.}
Prior work shows that CUDA memory bugs are highly exploitable~\cite{fun-profit,10.1145/3722041.3723099}. 
They can enable control-flow hijacking~\cite{di2016study,miele2016buffer}, mind-control attacks~\cite{park2021mind}, and side-channel attacks~\cite{10.1145/2801153,10.1145/3243734.3243831,zhao2024owl}, 
leading to severe inference degradation or arbitrary code execution. 
These findings motivate our systematic analysis of CUDA kernels to detect memory bugs in them.

\indentbolditalicparagraphnospace{Static techniques}
have been developed to analyze CUDA kernels. ProofWright~\cite{chatterjee2025proofwright} 
verifies kernels against specifications generated by LLMs. 
COVT~\cite{li2019detecting} translates CUDA into C/C++ 
and then detects integer and buffer overflows. 
Honeycomb’s validator~\cite{mai2023honeycomb} 
ensures memory accesses remain within predefined coarse-grained regions. 
Tong \etal{} extend Dartagnan with GPU consistency models~\cite{tong2024towards}.  
Methods based on symbolic execution 
have also been proposed~\cite{betts2012gpuverify,li2010symbolic,li2014practical,li2013se,li2012gklee,pereira2016verifying}.
Overall, these static techniques struggle with dynamically 
sized memory buffers, do not support the widely used \texttt{Tensor} library, and cannot handle varied kernel launch configurations.
We evaluate three of them~\cite{mai2023honeycomb,li2012gklee,pereira2016verifying} in Section~\ref{sec:coverage} and find they are limited in usability and less effective than \STool{}. Commercial static tools also exist~\cite{axivioncuda,parasoftcuda}, but their technical details are unavailable for analysis.

\indentbolditalicparagraphnospace{Dynamic Techniques.}
NVIDIA provides several mechanisms to enforce GPU memory safety. 
The GPU MMU reports ``illegal memory access'' 
errors when unmapped virtual addresses are accessed~\cite{windowsgpummu}. 
CUDA-MEMCHECK~\cite{cudamemcheck} and its successor Compute Sanitizer~\cite{computesanitizer} record metadata for GPU buffers, enabling detection of a broader range of memory errors at the cost of increased runtime overhead.

Researchers have also devoted efforts along this direction.
GPUShield~\cite{lee2022securing}, GPUArmor~\cite{ziad2025gpuarmor}, cuCatch~\cite{cuCatch}, and LAK~\cite{zhang2022lak} build on principles similar to Compute Sanitizer, with the first three emphasizing lower runtime overhead and the latter using lock-key metadata 
to provide stronger memory safety. 
GMOD~\cite{di2018gmod,di2020efficient} pads memory buffers and detects overflows by monitoring changes to the padding.
IMT~\cite{sullivan2023implicit} and LMI~\cite{lee2025let} ensure GPU memory safety with ECC 
and unused upper bits of pointers, respectively.
Race detection tools, including BARRACUDA~\cite{eizenberg_barracuda_2017}, GRace~\cite{grace}, GMRace~\cite{gmrace}, and LDetector~\cite{li2014ldetector}, 
track the last-accessing thread for each memory location and identify unsynchronized concurrent accesses. 
Despite these advances, these techniques rely on specific inputs to trigger bugs, require instrumentation or hardware support, and many incur large runtime overhead. 

Fuzzing techniques have also been proposed for CUDA kernels and machine learning
libraries~\cite{singh2026cufuzz, li2026cuda_fuzzing, liu2023nnsmith, ou2025mirrorfuzz}. 
However, they overlook kernel usage context within LLM inference systems
or target inference frameworks rather than CUDA kernels. 
They are ineffective in detecting memory bugs in CUDA kernels used in LLM inference systems.

\section{Conclusion}

This paper presents \Tool{}, an automated tool combining 
dynamic model profiling with CUDA-specialized symbolic execution to detect CUDA memory bugs in LLM inference systems. 
\Tool{} identifies 181 CUDA memory bugs in popular LLM inference systems when they perform inference with real models, demonstrating its practical effectiveness. Future work can improve 
the efficiency of symbolic execution for CUDA kernels 
and extend \Tool{} to support NPU kernels.

\section*{Acknowledgments}
We thank the anonymous reviewers and shepherd for their insightful feedback 
and valuable suggestions, and Yotta Labs for providing the compute resources.  
This work was supported in part by the U.S. National Science Foundation (NSF) under awards CNS-1955965 and CCF-2145394, 
and by the National Natural Science Foundation of China under grant 92582108.
Linhai Song is the corresponding author.

\bibliographystyle{ACM-Reference-Format}
\bibliography{cuda}

\end{document}